\documentclass[preprint,groupedaddress,floatfix]{revtex4}
\begin{document}
\title{The effect of the incident relative phase on the four-wave mixing field and the electromagnetically induced transparency}
\author{Yueping Niu, Shangqing Gong}
\affiliation{Key Laboratory for High Intensity Optics,Shanghai
Institute of Optics and Fine Mechanics, Shanghai 201800,
P.R.China}

\begin{abstract}
In a $\Lambda$-type system employing a two-photon pump field, a
four-wave mixing field can be generated simultaneously and hence a
closed-loop system forms. We study theoretically on the effect of
relative phase between the two incident fields on the generated
four-wave mixing field and the electromagnetically induced
transparency. It is found that the phase of the generated
four-wave mixing field is a sum of the incident relative phase and
a fixed phase which is irrelative to the incident relative phase.
Hence the total phase of the closed-loop system is independent of
the incident relative phase. As a result, the incident relative
phase has no effect on electromagnetically induced transparency,
which is different from the case of a $\Lambda$-type loop system
closed by a third incident field.

\vspace{0.5cm} PACS numbers: 42.50.Gy, 32.80.Qk, 42.50.Hz.
\end{abstract}

\maketitle
\section{Introduction}
  The influence of three laser fields of which two often jointly
couple to one dipole forbidden transition of a three-level atom
makes the generation and amplification of a fourth usually
high-frequency field possible. In this aspect, Harris et al. [1]
and Petch et al. [2] have studied the generation of four-wave
mixing (FWM) theoretically in detail. Dorman and Marangos [3, 4]
have realized high efficiency of generating four-wave mixing field
in krypton atom experimentally. Later, the effect of four-wave
mixing on electromagnetically induced transparency (EIT) has also
been studied and the asymmetry of EIT profile been demonstrated
[5, 6].

  The effect of electromagnetically induced transparency (EIT) [7]
has been the focus of a large number of theoretical and
experimental studies in the past few decades. In a simple
$\Lambda$-type system, which interacts with two incident fields,
the relative phase between the two incident fields will not
influence its properties. In a closed-loop $\Lambda$-type three
level system, however, the relative phase among the three incident
fields not only influences the temporary evolution of the system
[8], but also determines the values of the stationary populations
in the system [9]. Kosachiov et al. stressed that only at values
of general relative phase divisible by $\pi$, did EIT in the
system take place. At any other values, the excited level was
populated. Furthermore, with increasing of intensity of the
applied fields, the $\Lambda$-type closed-loop system became more
and more sensitive to the change of the incident relative phase.
In a double-$\Lambda$ scheme, Korsunsky et al. have studied the
phase-dependent nonlinear optics [10]. They found that at
resonance or near-resonance excitation of atoms, particular
relations for the field phases, amplitudes, and frequencies
ensured the coherent population trapping state. Experimentally,
phase-sensitive population dynamics was observed in three- [11,
12] and four-level [13, 14] closed-loop systems.

  In the present paper, we investigate the phase of the generated
FWM field and the effect of the incident relative phase on EIT in
a $\Lambda$-type three level system employing a two-photon pump
field [3-6]. The FWM field generates simultaneously and gains a
phase of $\theta+\phi$. Because the ground and the lower excited
states are closed by the generated FWM field, the total phase of
this system becomes $\phi$ and is independent of the incident
relative phase $\theta$. Consequently, the incident relative phase
does not influence the property of the closed-loop system.

\section{The model and the generated four-wave mixing field}

  We consider an atomic system consisting of a ground state $\vert c\rangle$, a
lower excited state $\vert b\rangle$ and an upper excited state
 $\vert a\rangle$ as shown in Fig.1. Since not all three transitions can be
dipole allowed, we here assume the $\vert a\rangle$- $\vert
c\rangle$ transition to be of two-photon nature and so that
$\omega_p$ is selected to be the two-photon pump field. The
four-wave mixing field is around the dipole allowed transition
$\vert b\rangle$- $\vert c\rangle$ at frequency $\omega_f$. The
Hamiltonian $H$ in the interaction picture describing the
interaction of this three level atom with three laser fields with
frequencies $\omega_d$, $\omega_p$ and $\omega_f$ can be cast in
the form

\begin{equation}
H(t)= \left(
            \begin{array}{ccc}
            \Delta_p&\Omega_d&\Omega_p\\
            \Omega_d^*&\Delta_f&\Omega_f\\
            \Omega_p^*&\Omega_f^*&0
            \end{array}
             \right),
\end{equation}
where $\Omega_i(i=p,d,f)$ refers to the Rabi frequency
corresponding to $\omega_i(i=p,d,f)$. We assume that
$\Omega_d=\Omega_{d0}\exp(-i\theta)$ and
$\Omega_d^*=\Omega_{d0}\exp(i\theta)$ while $\Omega_p^*=\Omega_p$.
$\theta$ represents the relative phase which is determined by the
phases of the two incident fields and the corresponding
transitions. $\Delta_i$ corresponds to the detunings given by
\begin{equation}
\Delta_p =\omega_{ac}-\omega_{2p},\Delta_d
=\omega_{ab}-\omega_d,\Delta_f =\omega_{bc}-\omega_f.
\end{equation}

Here, energy conservation $\Delta_p=\Delta_d+\Delta_f$ has been
assumed so that the above Hamiltonian becomes time-independent.
Thus, the system Liouville equation for the density matrix element
can be written as
\begin{equation}
\dot{\rho}(t)=-i[H(t),\rho]+\Lambda\rho,
\end{equation}
where $\rho$ is the density matrix in the interaction picture and
$\Lambda\rho$ represents the irreversible decay part in the
system. $\Lambda$ is a phenomenologically added decay terms
corresponding to all the incoherent processes. According to steady
state solution of Eq.(3), we can obtain the density matrix element
$\rho_{bc}$. Expanding it in power of $\Omega_f$, we get
\begin{equation}
\rho_{bc}=A+(B_1\Omega_f+B_2\Omega_f^*)+C|\Omega_f|^2+\ldots.
\end{equation}

  It is known that the response of a medium to an electric field is
governed by its polarization. Consider the coherent radiation
field generated by the four-wave mixing process, the polarization
can be written as:
\begin{equation}
P(\omega_f)=N(\mu_{bc}\rho_{cb}+\mu_{cb}\rho_{bc})=2N\mu_{cb}\rho_{bc},
\end{equation}
where $N$ denotes the number density of atom with dipole moment
$\mu_{cb}$. Because of
\begin{equation}
P_l(\omega_f)=\varepsilon_0\chi^1E_f,  and
P_{nl}(\omega_f)=\varepsilon_0\chi^3E_p^2E_d^*,
\end{equation}
we can get the expression of the FWM field according to the
treated methods reported by literature [2]:
\begin{equation}
E_f={\varepsilon_0\chi^3E_p^2E_d^*\over\chi^1}[exp({i\omega_f\chi^1z\over2c})-1].
\end{equation}
Here, phase-match condition is used.

\section{The effect of the incident relative phase on EIT}
  Solving the equation of motion of the density matrix elements in
steady state, we obtain the element $\rho_{bc}$. Though it is
complicated, we express it as a power expansion of the Rabi
frequency $\Omega_f$, just as shown in Eq.(4). Through simple
calculation, the relations between the expanding parameters and
the incident relative phase are achieved. A is proportional to
$exp(i\theta)$, $B_2$ is proportional to $exp(2i\theta)$ while
$B_1$ is independent of the relative phase $\theta$.

  Then, we display the quantity $\Omega_f$ as a function of the expanding parameters $A$, $B_1$ and
  $B_2$:
\begin{equation}
\Omega_f={AB_1^*-A^*B_2\over(B_1B_1^*-B_2B_2^*)}.
\end{equation}

  It is easy to point out that the Rabi frequency of the FWM field
is proportional to $exp(i\theta)$. Furthermore, because the
expanding parameters are complex, we can re-write it as
$(a+bi)exp(i\theta)$. Hence, for the generated FWM field, the
corresponding phase is $\theta+\phi$. $\phi$ is equal to
$\arctan(b/a)$, which is determined by the detunings, the Rabi
frequencies of the two incident fields and the decaying rates. As
the three quantities are determined, $\phi$ becomes a fixed value
and is completely independent of the incident relative phase.

  In the following, we focus on the effect of the relative phase
between the two incident fields on EIT. In this paper, we assume
that the relative phase $\theta$ is added to the coupling field,
i.e., $\Omega_d=\Omega_{d0}exp(-i\theta)$. At the same time, the
generated FWM field can be expressed as
$\Omega_f=\Omega_{f0}exp[-i(\theta+\phi)]$. Accordingly, the total
phase of the closed-loop system is $\phi$. As a result, the
incident relative phase has no effect on the population of the
upper level $\vert a\rangle$, i.e. EIT. We show this property in
Fig. 2. This phenomenon is completely different from that of the
$\Lambda$-type scheme closed by a third incident field. Just as
shown by Kosachiov et al., a radio-frequency field closed their
system and its EIT depended on the incident relative phase
crucially. Only at values of general relative phase divisible by
$\pi$, did EIT in the system take place. This is the greatest
difference between the two loop schemes closed by an incident
field and by a generated field respectively.

  In the following, we consider the case of multi-photon
resonance. Analytic expressions for $\Omega_f$ and $\rho_{aa}$
could be presented concisely. First of all, we discover the FWM
field for the stationary case:
\begin{equation}
\Omega_f={\Omega_d\Omega_p(\Omega_d^2+\Omega_p^2)\over\gamma(\Omega_d^2-\Omega_p^2)}e^{i(\theta-\pi/2)}.
\end{equation}

  It is clear that Eq. (9) fully shows the phase of the generated
FWM field. If the relative phase between the two incident fields
is $\theta$, the generated FWM field will gain a phase of
$\theta-\pi/2$. Hence, even if the incident relative phase equals
to zero, the generated FWM field still achieves a phase of
$-\pi/2$. Solving Eq. (3) under stationary condition, we could
obtain the population distribution of the upper excited level
$\vert a\rangle$:
\begin{equation}
\rho_{aa}={2\gamma^2\Omega_d^2\Omega_p^2(\Omega_d^2-\Omega_p^2)^2\over\gamma^4\Omega_p^6-\Omega_d^4\Omega_p^2(\gamma^4+4\gamma^2\Omega_p^2-4\Omega_p^4)+\Omega_d^6(\gamma^4+2\gamma^2\Omega_p^2+4\Omega_p^4)+\Omega_d^2\Omega_p^4(2\gamma^2\Omega_p^2-\gamma^4)}.
\end{equation}

  Apparently, now the population of the level $\vert a\rangle$ is independent of
the incident relative phase $\theta$. In particular, when
$\Omega_d=\Omega_p$, the upper level is unpopulated and EIT
appears. Therefore, it is the relative amplitude of the incident
fields but not the incident relative phase that control the EIT.
Kosachiov et al. has studied the $\Lambda$-type three level system
closed by a radio-frequency field and found that the population of
the upper level $\vert a\rangle$ depends not only on the intensity
of the fields but also on the relative field phase. By changing
the incident relative phase correspondingly, one can both destroy
and restore again the EIT. Comparatively, we can say that the
response of the loop $\Lambda$-type system closed by the generated
FWM field is completely different from the case closed by an
incident radio-frequency field.

\section{Conclusions}
  We investigated the phase of the generated four-wave mixing field
and the effect of the relative phase between the incident fields
on the EIT in a $\Lambda$-type three level system employing a
two-photon pump field. Because the phase of the simultaneously
generated four-wave mixing field is a sum of the incident relative
phase and a fixed phase, the total phase of the closed loop system
is independent of the incident relative phase. Therefore, it was
found that the occurrence of EIT did not lie on the incident
relative phase but on the relative amplitude of the two incident
fields. This property was different from the ordinary case closed
by a third incident field. In general, the behavior of the system
was very different in the case of loop interactions closed by a
generated FWM field.

\section*{Acknowledgement}
This work is supported by the National Natural Sciences Foundation
of China (Grant No. 10234030) and the Natural Science Foundation
of Shanghai (Grant No. 03ZR14102).

\vspace{1cm} Fig. 1 Loop $\Lambda$-type three level system closed
by the generated four-wave mixing field.

\vspace{1cm} Fig. 2 Population of the upper level $\vert a
\rangle$ (EIT) as functions of the detuning $\Delta_p$ and the
relative phase $\theta$. Parameters are $\Omega_d=\Omega_p=0.2$,
and $\gamma=0.5$.
\end{document}